\documentstyle[aps,twocolumn,epsf]{revtex}

\def \no{\noindent}

\def\thalf{{\textstyle{\frac{1}{2}}}}
\def\fourth{{\textstyle{\frac{1}{4}}}}
\def\dotp{\! \cdot \!}
\def\crossp{\! \times \!}

\begin{document}
\twocolumn[\hsize\textwidth\columnwidth\hsize\csname @twocolumnfalse\endcsname

\title{
The Quenching of the Axial Coupling in Nuclear and Neutron-Star Matter
}
\author{G.~W.~Carter and M.~Prakash}
\address{Department of Physics and Astronomy, SUNY at Stony Brook, Stony Brook,
NY 11794-3800}

\date{24 September 2001}

\maketitle

\begin{abstract}

Using a chirally invariant effective Lagrangian, we calculate the
density and isospin dependences of the in-medium axial coupling,
$g_A^*$, in spatially uniform matter present in core collapse
supernovae and neutron stars. The quenching of $g_A^*$ with density in
matter with different proton fractions is found to be
similar. However, our results suggest that the quenching of the
nucleon's $g_A^*$ in matter with hyperons is likely to be
significantly greater than in matter with nucleons only. \\

\no{PACS: 97.60.Jd, 21.65.+f, 12.39.Fe, 26.60.+c } \\
\end{abstract}
]



The accurately measured beta decay lifetime of the neutron in vacuum,
$n \rightarrow p+ e^-+ \bar\nu_e$, fixes the ratio of the axial and
vector couplings of the neutron to be $|g_A/g_V|= 1.2601\pm0.0025$
\cite{PDG}.  Studies of beta decays in nuclei, however, have long
suggested that a value of $|g_A/g_V| \simeq 1$ better fits the observed
systematics \cite{betas}.  Such a lower value also appears to be
consistent with pion-nucleus optical potentials \cite{piops} and the
systematics of Gamow-Teller resonances in nuclei \cite{gtr}.
Data from muon capture on nuclei, in which the relevant momentum transfer 
$q^2 \simeq -0.9m_\mu^2$, have been recently analyzed
including detailed nuclear structure effects \cite{muons} with the  
conclusion that a quenched $g_A$ (assuming
$g_V\cong 1$) is not necessary inasmuch as the vacuum value of $g_A$
adequately accounts for the data.  

The above experiments measure space-like axial transitions in nuclei.
At finite density, however, space-like and time-like axial matrix
elements are not necessarily equal, since Lorentz invariance is
broken. In fact, there are indications from experiments with
first-forbidden $\beta$ decays of light nuclei that the time-like
axial charge {\em increases} by about 25\% in medium \cite{wtb}.  A
theoretical expectation of this enhancement in terms of soft-pion
exchanges has been offered in Ref.~\cite{towner}.

The space-like quenching of $g_A$ in nuclei at low 
momentum transfers has been
attributed to a combination of effects, including the partial
restoration of chiral symmetry in a nuclear medium, the direct
participation of the $\Delta(1232)$ in renormalizing the in-medium
axial-vector current, and tensor correlations in nuclei in which shell
structure effects are important~\cite{piops}. An illuminating
discussion of the extent to which the quenching phenomenon is
intrinsic to the basic property of the ``vacuum'' defined by a
baryon-rich medium has been given by Rho (cf. Ref.~\cite{rho} and
references therein).  Later discussions of the quenching phenomenon in
chiral approaches to spatially uniform matter can be found in
Ref.~\cite{chiral}.  The issue of breaking and restoring fundamental
symmetries at large baryon density is presently intractable in lattice
gauge simulations. We therefore employ an effective field-theoretical
approach, based upon chiral symmetry, to consider medium modifications
of the nucleon's axial coupling.

The precise value of the in-meidum axial coupling, denoted by $g_A^*$
hereafter, in the dense matter encountered in astrophysical phenomena
such as core collapse supernovae and neutron
stars is crucially important.  In these cases,
weak interaction rates (that are $\propto g_A^{*2}$) drive the dynamics
from beginning to end.  In contrast to laboratory nuclei, in which the
weak processes occur at nuclear to subnuclear densities, 
astrophysical settings feature supernuclear densities.  This
highlights the need for knowledge of $g_A^*$ well beyond nuclear
densities and in uniform matter with varying isospin content.  In the
supernova environment, $e^-$-capture reactions on neutrons and nuclei
begin the process of neutronization and decrease of the total lepton
fraction, $Y_L=n_L/n_B$, whose value after $\nu$-trapping ($\simeq
0.38-0.4$) determines the masses of the homologous core and initial
proto-neutron star and thus the available energy for the shock and
subsequent $\nu$-emission \cite{Bur00}.  The bremsstrahlung $(n+n
\rightarrow n+n+\nu+\bar\nu)$ and modified Urca ($n+p \rightarrow
n+n+e^++\nu+\bar\nu$) processes dominate in the production and
thermalization of $\mu$ and $\tau$ neutrinos \cite{TBH00}.  The
$\nu$-luminosity and the time scale over which $\nu$s remain
observable from a proto-neutron star are also governed by charged
and neutral current interactions involving baryons at high density
\cite{PNSs}.  The long-term cooling of a neutron star, up to a 
million years of age, is controlled by $\nu$-emissivities from the
densest parts of the star; thereafter the star is observable through
photon emissions, which may allow us to determine the star's mass,
radius, and internal constitution \cite{cool}.

In this Letter, we focus on the behavior of the in-medium space-like $g_A^*$,
both with increasing density and with varying isospin content, in
spatially uniform matter in which shell effects characteristic of
finite nuclei are absent. This enables us to expose cleanly the role
played by chiral symmetry alone.  Our results are directly relevant to
astrophysical situations in which spatially homogeneous matter are
encountered. We consider the cases of isospin symmetric matter (proton
fraction $x=n_p/n_B=1/2)$, pure neutron matter ($x=0$), and
neutron-star matter in which the equilibrium value of the proton
fraction, $\tilde x$, is determined by the conditions of beta stability
and charge neutrality. The latter two cases have been largely ignored
in the literature, but are essential for astrophysical modeling in
which neutrinos play a dominant role. 

The nucleon coupling to the axial current is directly determined through the 
matrix element
\begin{eqnarray}
\langle N(p_2) \vert \vec{A}_\mu \vert N(p_1) \rangle =
\bar{u}(p_2) \thalf\vec{\tau}\left[ \gamma_\mu\gamma_5 g_A \right.&& 
\nonumber \\
&&\hspace{-2cm} \left. + q_\mu \gamma_5 h_A(q^2)
\right] u(p_1) \,, 
\label{axialdecomp}
\end{eqnarray}
where the $\bar{u}(p_2)$ and $u(p_1)$ are the spinor solutions of the
Dirac equation for nucleons and $q=p_2-p_1$.  In this form, $g_A$ is
the axial coupling and $h_A(q^2)$ the form factor.  Taking medium
effects into account does not change this definition when we consider
a medium-modified $g_A^*$.  In the applications of interest here,
momentum and energy transfers are moderately low. We therefore
concentrate on medium modifications of $g_A$~\cite{note}.  
The axial current can be computed from first principles. Explicitly,
\begin{equation}
A_\mu^a = - \sum_i \frac{\delta {\cal L}}{\delta \partial^\mu \Phi_i}
\,\delta^a_5 \Phi_i \,,
\label{axialdef}
\end{equation}
where the sum runs over all fields in the Lagrangian and $\delta_5^a
\Phi_i$ is the change in each field under an axial transformation.
The fact that the axial coupling is related to pion dynamics requires
that our approach should involve the spontaneous breaking of chiral
symmetry.  We therefore use a chiral Lagrangian to describe the
interactions between baryons, keeping in mind that it should
reproduce the vacuum value of $g_A$, equilibirum empirical
properties of isospin symmetric and asymmetric matter, reasonably
describe closed-shell nuclei, and account for observables in
pion-nucleon interactions. Fortunately, it is possible to achieve
these objectives with mean fields computed at the tree level.
Originally motivated by the gluonic trace anomaly in QCD, a model
incorporating a heavy glueball field was devised which replaced the
chiral ``sombrero'' potential with a logarithmic one
\cite{GBs,note1}.  For our purposes, it will be sufficient to ``freeze''
the glueball field at its vacuum value, which shall have negligible
consequences since the gluon condensate changes only slightly at
finite density \cite{hre94}.  Specifically, we have
\begin{eqnarray}
{\cal L} &=& 
\thalf\Delta_\mu\sigma\Delta^\mu\sigma
+ \thalf\Delta_\mu\vec{\pi}\dotp\Delta^\mu\vec{\pi}
- U(\sigma,\vec{\pi})
\nonumber \\
&-& \fourth\omega_{\mu\nu}\omega^{\mu\nu}+\thalf m_\omega^2\omega_\mu\omega^\mu
 - \fourth\vec{F}_{\mu\nu}\vec{F}^{\mu\nu} 
 + \thalf m_\rho^2\vec{\rho}_\mu \dotp \vec{\rho}^\mu
\nonumber\\
&-& \fourth\vec{G}_{\mu\nu}\vec{G}^{\mu\nu} 
+ \thalf \left( m_\rho^2 
   + \frac{m_a^2-m_\rho^2}{\sigma_0^2}\sigma^2\right)
  \vec{a}_\mu \dotp \vec{a}^\mu
\nonumber\\
&+& G^4 \left(\omega_\mu\omega^\mu\right)^2
+ \bar{N}\left[ 
\gamma^\mu\left(i\partial_\mu-g_\omega\omega_\mu\right) \right. 
\nonumber \\
&-& \left. g\left(\sigma + i \vec{\tau}\dotp\vec{\pi}\gamma_5\right)
+ \thalf g_\rho\gamma^\mu\vec{\tau}\dotp\left(\vec{\rho}_\mu-
  \vec{a}_\mu\gamma_5\right) \right] N 
\nonumber\\[1.5mm]
&+& \thalf D\bar{N}\gamma^\mu\vec{\tau}\dotp\left[
\vec{\pi}\crossp\Delta_\mu\vec{\pi} + \gamma_5 \left( \sigma\Delta_\mu
\vec{\pi} - \vec{\pi}\Delta_\mu\sigma\right)\right] N 
\,,
\nonumber\\[2mm]
U(\sigma&\hspace{-10pt},&\hspace{-10pt}\pi) = \thalf B \left[ 
\frac{\sigma^2+\vec{\pi}^2}{\sigma_0^2} 
- 1
- \ln\left(\frac{\sigma^2+\vec{\pi}^2}{\sigma_0^2} \right) 
\right] \,.
\label{lagran}
\end{eqnarray}
The field strength tensors are 
\begin{eqnarray}
\vec{F}_{\mu\nu} &=& \partial_\mu\vec{\rho}_\nu-\partial_\nu\vec{\rho}_\mu 
+ g_\rho \vec{\rho}_\mu\crossp\vec{\rho}_\nu 
+ g_\rho \vec{a}_\mu\crossp\vec{a}_\nu\,, \nonumber\\
\vec{G}_{\mu\nu} &=& \partial_\mu\vec{a}_\nu-\partial_\nu\vec{a}_\mu 
+ g_\rho \vec{\rho}_\mu\crossp\vec{a}_\nu 
- g_\rho \vec{\rho}_\nu\crossp\vec{a}_\mu\,, \nonumber
\end{eqnarray}
with the covariant derivatives 
\begin{eqnarray}
\Delta_\mu\sigma &=& \partial_\mu\sigma + g_\rho\vec{a}_\mu\dotp\vec{\pi}\,,
\nonumber\\
\Delta_\mu\vec{\pi} &=& \partial_\mu\vec{\pi}
+ g_\rho\vec{\rho}_\mu\crossp\vec{\pi}-g_\rho\sigma\vec{a}_\mu  \,.
\nonumber
\end{eqnarray}
The $N$ are isospinor nucleons, $\sigma$ and $\pi$ are scalar and
pseudoscalar isovector chiral mesons, and $\rho$ and $a_1$ are
isovector mesons which are the vector and psuedovector representations
of a gauged chiral symmetry group and make vector meson dominance
inherent in the model \cite{thesis}.  The $\omega$ is a chiral singlet
vector field which supplies the short-ranged repulsion necessary for
saturation, with a self-interaction term included to soften nuclear
matter.  The non-standard coupling between the nucleon and meson
fields, with the coefficient $D$, serves to generate the physical value
of $g_A$ at the mean-field level while respecting chiral symmetry
\cite{lee}. Note that the pion-nucleon interactions in Eq.~(\ref{lagran})
are chiral invariant.

The effective chiral Lagrangian of Eq.~(\ref{lagran}), like that of the
Walecka model and its variant approaches to nuclear matter
\cite{SW86}, is to be used at the mean-field level.  This model is
able to describe closed-shell nuclei similar in accuracy to that
obtained in Quantum Hadrodynamical models \cite{hre94}.  Calculations
of low-energy pion-nucleon scattering also compare well with data
\cite{chiral}.  The scaling of the physical constants, effective
masses, and the quenching of $g_A$ in isospin symmetric matter are
also detailed in Ref.~\cite{chiral}.  Note also that the
Goldberger-Treiman relation, $g_{\pi NN} f_\pi = g_A M$, remains valid
at finite baryon density, since the model encodes chiral symmetry.

We turn now to spatially uniform matter with an unequal number
of neutrons ($n$) and protons ($p$).  In medium, the $\sigma$,
$\omega_0$, and $\rho_0^3$ fields develop non-zero expectation values,
which we denote by $\bar\sigma$, $\bar\omega$, and $\bar\rho$.  These
are obtained by solving the equations of motion:
\begin{eqnarray}
\frac{B}{\bar\sigma}\left(1-\frac{\bar\sigma^2}{\sigma_0^2}\right) &=&
g\langle\bar N N \rangle = g\left(n^s_p + n^s_n\right) \,,
\nonumber\\
m_\omega^2 \bar\omega + 4 G^4 \bar\omega^3 &=& 
g_\omega\langle N^\dagger N \rangle = g_\omega \left(n_p+n_n\right) \,,
\nonumber\\
m_\rho^2 \bar\rho &=& 
g_\rho\langle N^\dagger \tau^3 N \rangle = g_\rho\left(n_p-n_n\right) \,.
\label{eom}
\end{eqnarray}
Above, $n_i^s$ and $n_i$ with $i=n,p$ denote the scalar and baryon
number densities, and $\tau^3$ is the third component of the isospin
operator.  

From the Dirac equation,
\begin{equation}
\left( i\gamma^\mu\partial_\mu - g_\omega\gamma^0\bar\omega 
- \thalf g_\rho \tau^3 \gamma^0 \bar\rho - g\bar\sigma \right) N = 0,
\label{diraceq}
\end{equation}
we find the effective mass, $M^* = g\bar\sigma$, and the effective
chemical potentials,
$\mu_i^* = \mu_i - g_\omega \omega_0 - \thalf g_\rho \tau^3 \rho_0^3$.
The scalar and baryon densities 
depend self-consistently on these effective parameters: 
\begin{eqnarray}
n^s_i &=& \frac{M^*}{2\pi^2}\left[k_{f_i}E_{f_i}^* 
-M^{*\,2}\ln\left(\frac{k_{f_i}+E_{f_i}^*}{M^*}\right)\right]
\,, \nonumber \\
n_i &=& \frac{k_{f_i}^3}{3\pi^2}  \,,
\end{eqnarray}
where $E_{f_i}^*= \sqrt{k_{f_i}^2+M^{*\,2}}$ and 
$k_{f_i} = \sqrt{\mu^{*\,2}_i-M^{*\,2}}$.
In pure neutron matter, $n_p=0$ and $n_p^s=0$. 
In beta-stable and charge neutral neutron-star matter, the constraints 
\begin{eqnarray}
\mu_n - \mu_p = \mu_e &=& \mu_\mu \,,
\nonumber \\
n_p - n_e - n_\mu &=& 0 \,, 
\label{constrs}
\end{eqnarray}
where $n_e$ and $n_\mu$ are the number densities of electrons and muons,
determine the equilibrium proton fraction.  With the 
solutions to Eqs.~(\ref{eom}) the energy density of matter,
\begin{eqnarray}
{\cal E} &=&
\thalf m_\omega^2\bar\omega^2 + G^4\bar\omega^4 + \thalf m_\rho^2\bar\rho^2 + 
\thalf B\left( \frac{\bar\sigma^2}{\sigma_0^2} - \ln\frac{\bar\sigma^2}{
\sigma_0^2} - 1 \right) \nonumber \\
&&
+ \sum_{i=p,n} \frac{1}{\pi^2}\int_0^{k_{f_i}} dk\,k^2\sqrt{k^2+M^{*\,2}}
\,,
\label{edens}
\end{eqnarray}
and the energy per baryon, $E/A = {\cal E}/n_B - M$, are easily computed.

In order to compute the axial current, we will need only the terms in
Eq.~(\ref{axialdef}) which are finite for a nonzero sigma mean field and
contribute to the nucleon matrix element Eq.~(\ref{axialdecomp}).  The
relevant terms in the Lagrangian, Eq.~(\ref{lagran}), will be the nucleon
kinetic energy and derivative coupling to the pion, the latter since
$\delta_5^a \pi^b = \delta^{ab}\sigma$.  Renormalization of $\pi-a_1$
mixing introduces the physical fields
\begin{equation}
\vec{\pi}' = \left(1-\frac{g_\rho^2\sigma^2}{m_a^{*2}}\right) \vec{\pi}
\, , \qquad
\vec{a}_\mu' = \vec{a}_\mu - \frac{g_\rho\sigma}{m_a^{*2}} 
\partial_\mu\vec{\pi} \,,
\label{pia1mix}
\end{equation}
where the effective mass of the $a_1$ meson depends on the sigma mean field as 
$m_{a}^{*2} = m_\rho^2 + (m_{a}^2-m_\rho^2)\bar\sigma^2/\sigma_0^2$.
This replacement leads to additional derivative terms from the $a_1$-nucleon 
coupling and, in terms of the redefined pion, contributions to the matrix 
element arise from
\begin{eqnarray}
{\cal L} &=& \bar N i \gamma^\mu\partial_\mu N \nonumber \\
&+& \bar N \gamma^\mu \gamma_5 
\left( \frac{Z_\pi^*-1}{2\sigma\sqrt{Z_\pi^*}} +
\frac{D\sigma\sqrt{Z_\pi^*}}{2\sigma_0^2} \right)
\vec{\tau}\dotp\partial_\mu\vec{\pi} N + \dots,
\end{eqnarray}
where the medium-dependent renormalization constant $Z_\pi^* = 1 -
g_\rho^2\sigma^2/m_a^{*2}$.  Using Eq.~(\ref{axialdef}), we find
\begin{equation}
\vec{A}_\mu = -
Z_\pi^*
\left( 1 + \frac{D\sigma^2}{2\sigma_0^2} \right) 
\bar N \frac{\vec{\tau}}{2} \gamma_\mu\gamma_5 N \,.
\label{axialcurrent}
\end{equation}
   Inserting this into
Eq.~(\ref{axialdecomp}) we have, in terms of the sigma mean field and
vacuum constants,
\begin{eqnarray}
g_A^* = \left( 1 + D\frac{\bar\sigma^2}{\sigma_0^2} \right)
\left( 1 - \frac{ g_\rho^2 \bar\sigma^2 }
  { m_\rho^2+\left(m_a^2-m_\rho^2\right)\bar\sigma^2/\sigma_0^2} \right) .
\label{ga}
\end{eqnarray}
Above, the first factor contains the ``bare'' axial coupling,
including the standard axial interaction and the mean-field dependent
modification from the $D$ term.  The second factor is the pion
renormalization constant of Eq.~(\ref{pia1mix}) with its mean-field
dependence made explicit.  Density dependence is thus generated by the
scalar mean field, $\bar\sigma$, alone.  We note that the functional
form of Eq.~(\ref{ga}), orginally derived in Ref.~\cite{chiral} for
isospin symmetric matter, remains intact for isospin asymmetric
matter. The numerical value of $g_A^*$, however, is controlled by the
magnitude of $\bar\sigma$ which depends on the proton fraction of
matter.  This relation and its implications to be discussed below are
among the prinicpal results of this work.

We now specify the physical parameters 
in Eq.~(\ref{axialdecomp})
and discuss quantitative results in matter.  Fixing the vacuum
value of the nucleon mass determines the chiral coupling $g = 9.2$. We
take $\sigma_0=f_\pi/\sqrt{Z_\pi}=102$ MeV, where $Z_\pi$ is the pion
renormalization constant necessitated by $\pi - a_1$ mixing.  The
value $D = 1.17$ reproduces $g_A = 1.26$ in vacuum.  The saturation of
nuclear matter at its empirical density $n_0\cong 0.15$ fm$^{-3}$ and
energy per baryon $E/A=-16$ MeV require $B = \left(323~{\rm
MeV}\right)^4$ and $g_\omega = 11.4$.  The omega self-interaction
strength is $G=0.19 g_\omega$, fixed to produce maximal softening of
the EOS while not generating spurious field solutions.  At saturation,
the resulting effective mass is $M^*=0.68M$ and the compression
modulus $K \cong 320$ MeV.  The rho coupling $g_\rho = 8.0$ yields the
empirical symmetry energy of $\cong $30 MeV. 

The procedure adopted above to fix the various couplings in
Eq.~(\ref{axialdecomp}) implies that the behavior of $g_A^*$ with
increasing density, including its value at the equilibrium density
$n_0$ of symmetric nuclear matter, is to be regarded as a prediction,
albeit within the confines of the model adopted for matter.  The EOS
of neutron-star matter at supra-nuclear densities is subject to the
constraint that it must support at least 1.44M$_\odot$, which is the
most accurately measured mass of the neutron star in the binary pulsar
PSR 1913+16~\cite{nobel}. Presently, more severe constraints at high
density are not available.

\begin{figure}[htb]
\centerline{{\epsfxsize=3.5in \epsfbox{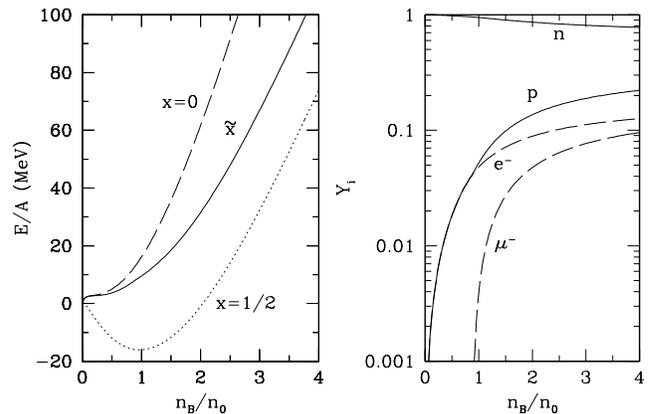}} }
\vspace*{-3cm}
\caption{Left panel: The energy per baryon in nuclear (dotted lines),
pure neutron (dashed) and beta-stable neutron-star matter (solid) as
functions of baryon density in units of $n_0=0.15~\rm{fm}^{-3}$. 
Right panel: Particle concentrations in beta-stable neutron
star matter.}
\label{ga_fig1}
\end{figure}

The left panel of Fig.~\ref{ga_fig1} shows how $E/A$ varies with
$n_B/n_0$ as the proton fraction $x$ is varied from nuclear to
beta-stable neutron-star to the ideal case of pure neutron-star
matter. The right panel shows the concentrations $Y_i=n_i/n_B$ of
$n$, $p$, $e^-$, and $\mu^-$ in neutron-star matter. The corresponding EOS
yields a maximum mass of $M_{max}/{\rm M}_\odot=1.6$; for this
configuration the central density $n_c/n_0\cong 7.4$ and radius
$R=10.24$~km. These values are to be compared with $M_{max}/{\rm
M}_\odot=2.2$, $n_c/n_0\cong 5.3$, and $R=12.74$~km for the case of
pure neutron matter, which highlights the role of the isospin content
in matter for this EOS.

\begin{figure}[htb]
\centerline{{\epsfxsize=3.5in \epsfbox{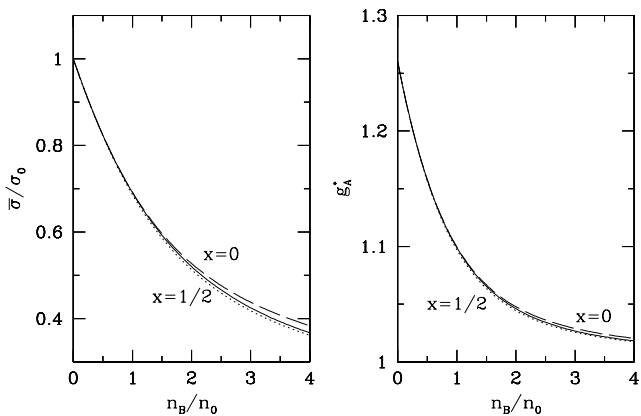}} }
\vspace*{-3cm}
\caption{The sigma mean fields (left panel) and the axial coupling
constants (right panel) in nuclear (dotted lines), pure neutron
(dashed), and beta-stable neutron-star matter (solid) as functions of
baryon density in units of $n_0=0.15~{\rm fm}^{-3}$. }
\label{ga_fig2}
\end{figure}

Fig.~\ref{ga_fig2} shows the density and proton fraction dependences
of the scalar mean field $\bar\sigma/\sigma_0$ (left panel) and the
axial-vector coupling $g_A^*$ (right panel).  The quenching of both
of these quantities with increasing $n_B$ is clearly evident, with 
a depreciation of 12\% at $n_0$ and a 19\% drop at $4n_0$.  {\em The
relatively mild variation with $x$ is important insofar as guidance
for the quenching of $g_A$ from laboratory studies (which sample a
narrow range in $x$) of nuclei have the potential of being directly
and immediately useful in astrophysical applications (in which a
broader range of $x$ is sampled).}
For modeling purposes, a simple parametrization of $g_A^*$ in terms of 
baryon density is
\begin{equation}
g_A^* \simeq g_A \left( 1 - \frac{n_B}{4.15 \left(n_0+n_B\right)} 
\right)\,.
\end{equation}
This expression matches the results of Eq.~(\ref{ga}) to within 1\% 
accuracy for all $n_B \le 4.5 n_0$.

The quenching of $g_A$ considered in this work stems chiefly from the
medium-dependent scalar field and $\pi-a_1$ mixing.  The combination
of these two effects has been shown to be virtually independent of the
isospin content, suggesting that the in-medium behavior observed in
nearly iso-symmetric matter will be present in neutron and stellar
matter as well.
The excitation of
the $\Delta$, which is crucial in non-relativistic descriptions of
quenching, has yet to be satisfactorily implemented in a
relativistic field theory. 
Its addition would likely lead to further reduction.

It is also worthwhile to point out here that $\bar\sigma$ falls more
rapidly with density in matter with hyperons than without hyperons
(see Fig.~6 of Ref.~\cite{KPE}). This is mainly due to the presence of
additional baryonic components with dissimilar masses.  Consequently,
the nucleon's $g_A^*$ would be quenched to a greater extent in the
presence of hyperons.  A verification of this expectation would
require an extension of flavor symmetry, which has been attempted
recently with only limited success \cite{psssg}.  Our results in this
work suggest that the extension of the chiral model to incorporate
strange and Delta resonances with the full effects of relativity would
be worthwhile.

We thank Gerry Brown, Ernest Henley, and Paul Ellis for helpful discussions.
This work was supported by the US-DOE grant DE-FG02-88ER40388.


\begin{thebibliography}{99}
\bibitem{note1} 
The Lagrangian we use here was first formulated in the
spirit of the Walecka model for nuclear matter and finite nuclei
\cite{hre94}.  Although variants of the Walecka model that include
nonlinear interactions have been very successful phenomenologically,
they do not exhibit the spontaneous breaking of chiral symmetry, which
seems to characterize the hadronic vacuum. They therefore lack a
natural description of $g_A$.  On the other hand, the linear sigma
model of chiral symmetry broken with a ``sombrero'' potential
leads to an inadequate description of equilibrium nuclear matter and a
vacuum $g_A$ of unity.
\end{thebibliography}
\end{document}